\documentclass[aps,prl,twocolumn,groupedaddress,
showpacs,amsmath,amssymb]{revtex4}

\usepackage{graphicx}
\usepackage{dcolumn}
\usepackage{bm}

\begin{document}

\title{Peculiar Ferrimagnetism Associated with Charge Order
in Layered Perovskite GdBaMn$_{2}$O$_{5.0}$}

\author{A. A. Taskin}
\author{Yoichi Ando}

\affiliation{Central Research Institute of Electric Power Industry,
Komae, Tokyo 201-8511, Japan}


\begin{abstract}

The magnetic properties of GdBaMn$_{2}$O$_{5.0}$, which exhibits
charge ordering, are studied from 2 to 400 K using single crystals.
In a small magnetic field applied along the easy axis,
the magnetization $M$ shows a temperature-induced reversal which is
sometimes found in ferrimagnets. In a large magnetic field, on the other
hand, a sharp change in the slope of $M(T)$ coming from an unusual
turnabout of the magnetization of the Mn sublattices is observed. Those
observations are essentially explained by a molecular field theory which
highlights the role of delicate magnetic interactions between Gd$^{3+}$
ions and the antiferromagnetically coupled Mn$^{2+}$/Mn$^{3+}$
sublattices. 

\end{abstract}



\maketitle

Ferrimagnetism is a complex but intriguing type of magnetic ordering. It
occurs when antiferromagnetically aligned spins have different local
moments, resulting in their incomplete cancellation. The characteristics
of this type of ordering stems from the fact that it combines features
of both ferromagnetic (FM) and antiferromagnetic (AF) systems. Moreover,
if more than two spin sublattices are involved, a new level of
complexity, and hence new physics, can emerge \cite{Neel}. So far, only
a few families of ferrimagnetics with three magnetic sublattices --- spinel
ferrites and rare-earth garnets being the most famous examples
\cite{Belov} --- have been discovered. These materials are proved to be not
only fundamentally interesting, but also technologically important
\cite{applications}, inspiring the search for new promising magnetic
compounds.

Recently, half-doped $A$-site ordered manganite perovskites
$R$BaMn$_2$O$_{5+x}$ (where $R$ is a rare-earth element) have been
synthesized in an effort to clarify the role of the random potential
effect in the colossal magnetoresistance (CMR) phenomena
\cite{MillangeY,MillangeLa,Akahoshi,Ueda,Trukhanov}. It has been found
that in addition to the interesting electronic properties, these
compounds possess another potentially useful quality, a variability of
the oxygen content \cite{MillangeY,MillangeLa,Trukhanov,OxyDiff}. By
varying the oxygen concentration from $x$=0 to $x$=1, one can readily
change the valence state of Mn ions from 2+ through 3+ to 4+, generating
a variety of possible magnetic states. In particular, Mn ions adopt two
valence states, Mn$^{2+}$ and Mn$^{3+}$, and develop a charge order in
the reduced composition $R$BaMn$_2$O$_{5.0}$ ($x$=0)
\cite{MillangeY,MillangeLa}. Since most rare-earth ions ($R^{3+}$) are
also magnetic, $R$BaMn$_2$O$_{5.0}$ can be a three-sublattice magnetic
system with potentially ferrimagnetic type of ordering; among the
rare-earths, Gd with the largest spin and zero orbital angular momentum
would make a benchmark compound of this family. It is worth noting that
the layered crystal structure of these compounds naturally implies an
anisotropy in its magnetic properties and, therefore, single crystals
would have a great advantage for studying their magnetic behavior;
however, previous studies on $R$BaMn$_2$O$_{5.0}$
\cite{MillangeY,MillangeLa,Trukhanov} only used polycrystalline samples.

In this Letter, we present the first study of the magnetic properties of
GdBaMn$_{2}$O$_{5.0}$ single crystals, which show unusual behavior: At
$T_N$=144 K, a long-range order of Mn$^{2+}$ and Mn$^{3+}$ magnetic
moments is established. Upon decreasing temperature in a small magnetic
field along the $c$ axis, a magnetization reversal occurs at the
compensation point $T_{\rm comp}$, below which the net magnetic moment
is opposite to the magnetic-field direction. On the other hand, in a
large magnetic field the magnetization $M$ is positive at all
temperature, and shows a sharp kink in the $M(T)$ curve at $T_{\rm
comp}$. We show that the observed magnetic behavior is essentially
understood if Gd spins, which remain paramagnetic through $T_N$ and
$T_{\rm comp}$, are weakly coupled ferromagnetically
(antiferromagnetically) with Mn$^{3+}$ (Mn$^{2+}$) neighbors and
gradually align their spins parallel (antiparallel) to the Mn$^{3+}$
(Mn$^{2+}$) sublattice with decreasing $T$ in low fields. What is
special here is that in high magnetic fields, due to the weak Gd-Mn
coupling, the Gd spins are aligned along the external magnetic field,
which eliminates the magnetization reversal; intriguingly, in this
situation an abrupt turnabout of the magnetization of the Mn sublattices
occurs at $T_{\rm comp}$, which, to our knowledge, has never been
observed in any ferrimagnets. Hence, the peculiar ferrimagnetism in
GdBaMn$_{2}$O$_{5.0}$ is quite distinct from other transition-metal
oxides showing magnetization reversals \cite{Neel,Belov,TIMR,LaGdMnO3}.

The high-quality single crystals of GdBaMn$_{2}$O$_{5+x}$ used for this
study were grown by the floating-zone technique. The as-grown crystals
were annealed in flowing argon-hydrogen mixture at 600$^{\circ}$C for
several days to obtain the stoichiometry of $x$ = 0, which is confirmed
by the thermogravimetric analysis. Parallelepiped samples were cut and
polished with all faces adjusted to the crystallographic planes to
within 1$^{\circ}$. Magnetization measurements were carried out using a
SQUID magnetometer at fields up to 70 kOe applied parallel or
perpendicular to the $c$ axis.

\begin{figure}
\includegraphics*[width=18pc]{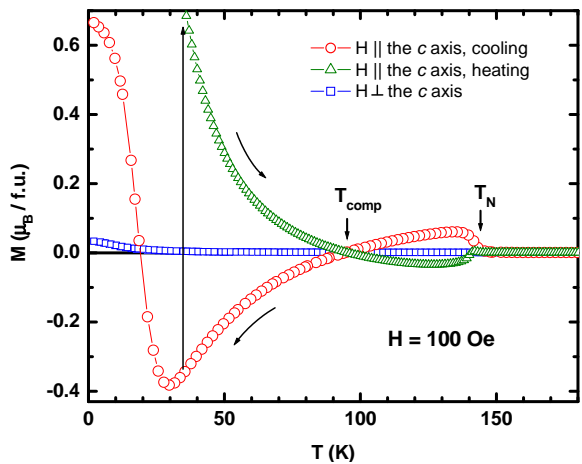}
\caption{(Color online) 
Temperature dependences of $M$ under different conditions. Upon cooling
in $H$ = 100 Oe applied along the $c$ axis (circles), $M(T)$ becomes
negative at $T_{\rm comp}$ and eventually returns to positive at lower
temperature; upon heating, nearly a ``mirror image" is observed after
applying a high magnetic field ($H$ = 3 kOe at $T$ = 35 K for the curve
shown by triangles). The squares show $M(T)$ measured in $H$ = 100 Oe
applied along the $ab$ plane for both cooling and heating.}
\label{fig1}
\end{figure}

Figure 1 shows the temperature dependences of the magnetization of
GdBaMn$_{2}$O$_{5.0}$ in a magnetic field of 100 Oe applied parallel or
perpendicular to the spin easy axis. Upon cooling in a magnetic field
along the $c$ axis, $M(T)$ (shown by circles) rapidly increases below
the N\'eel temperature, $T_N$=144 K, indicating the onset of a magnetic
ordering. About 10 K below $T_N$, $M(T)$ reaches a maximum, then starts
to decrease and eventually goes to zero at $T_{\rm comp}$=95 K. This
temperature, which is called the compensation point, marks the state
where magnetic contributions of all sublattices cancel each other. Below
$T_{\rm comp}$, $M(T)$ becomes negative, indicating that the net
magnetic moment is opposite to the external magnetic-field direction.
This phenomenon is called the {\it temperature-induced magnetization
reversal} \cite{Neel,Belov,TIMR,LaGdMnO3}. 

Upon further decrease in temperature and concomitant development of
sublattice magnetizations, the net magnetic moment grows too large to
remain in this metastable state. Eventually a magnetic field (even as
small as 100 Oe) can overcome the coercive force, leading to a rotation
of the net magnetization. As a consequence, another sign change of
$M(T)$ occurs at around 20--30 K. The effect of this rotation of the net
magnetic moment can be illustrated even more clearly in the following
experiment: First, the direction of the sublattice magnetizations in the
sample cooled down to an intermediate temperature in 100 Oe are switched
by applying a large magnetic field (for example, $H$ = 3 kOe at $T$ = 35
K as shown in Fig. 1). Then, the magnetic field is reduced down to 100
Oe again and $M(T)$ is measured upon heating. As can be seen in Fig. 1,
$M(T)$ shows almost a ``mirror image" of the magnetization recorded upon
cooling, indicating that the 180$^{\circ}$-spin-rotated state is kept
intact up to $T_N$ once it is formed.

A comparison of $M(T)$ measured along different crystallographic axes
reveals another important feature of the magnetic behavior in
GdBaMn$_{2}$O$_{5}$: The magnetization is strongly anisotropic with an
easy direction along the $c$ axis (see Fig 1), simplifying significantly
a possible arrangement of Mn$^{3+}$, Mn$^{2+}$, Gd$^{3+}$ spins.

In order to obtain further insight into the observed phenomena, the
field dependence of the magnetization has been measured at 2 K where
the magnetizations of all sublattices are expected to be fully
developed. As shown in Fig. 2(a), a magnetic field of about 1 kOe
applied along the $c$ axis is enough to reach the saturation. A much
higher magnetic field (about 50 kOe) must be applied perpendicular to
the easy axis to reach the same saturation level. The equality of the
saturation magnetization $M_{sat} \approx$ 5.85 $\mu_B$/f.u. achieved
along different crystallographic axes suggests that the obtained value
reflects the net saturated moment, which should be simply composed of
the sublattice magnetizations.

\begin{figure}[b]
\includegraphics*[width=19pc]{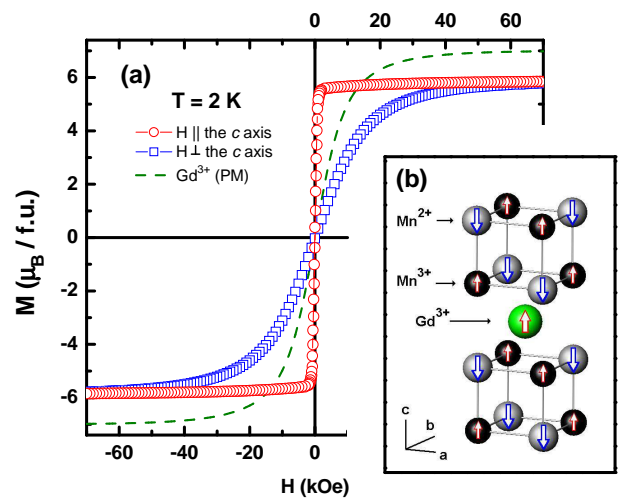}
\caption{(Color online) 
(a) $M(H)$ curves of GdBaMn$_{2}$O$_{5.0}$ with $H$ along the $c$ axis
(circles) and the $ab$ plane (squares). The calculated paramagnetic
response from the Gd$^{3+}$ sublattice is shown by the dashed line. (b)
A sketch of the crystal and magnetic structure of GdBaMn$_{2}$O$_{5.0}$,
showing three magnetic sublattices composed of Mn$^{2+}$, Mn$^{3+}$, and
Gd$^{3+}$. Nonmagnetic Ba and O are omitted for the sake of clarity.}
\label{fig2}
\end{figure}

Obviously, there is a unique combination of spin-only moments of the
three spin sublattices --- Mn$^{3+}$ ($S_{\rm Mn^{3+}}$=2), Mn$^{2+}$
($S_{\rm Mn^{2+}}$=5/2), and Gd$^{3+}$ ($S_{\rm Gd^{3+}}$=7/2) --- that
can be consistent with the experimentally observed value of the
saturated magnetization:
\begin{equation} 
g \mu_B \left ( \: S_{\rm Mn^{3+}}-S_{\rm Mn^{2+}}+S_{\rm Gd^{3+}}
\: \right ) = 6 \mu_B, 
\end{equation} 
where $g$=2 is the Land\'e $g$-factor and $\mu_B$ is the Bohr magneton.
This equation suggests an AF coupling between Mn$^{2+}$ and Mn$^{3+}$
and a FM (AF) interaction of Gd$^{3+}$ ions with the Mn$^{3+}$
(Mn$^{2+}$) sublattice. Note that the paramagnetic (PM) contribution of
Gd$^{3+}$ alone [shown by the dashed line in Fig 2(a)] would give a
larger saturated magnetization than observed. The spin-only magnetic
moments of Mn$^{3+}$, Mn$^{2+}$, and Gd$^{3+}$ ions are expected to be a
good approximation in GdBaMn$_{2}$O$_{5.0}$, because the high-spin (HS)
Gd$^{3+}$ ($4f^7$) and Mn$^{2+}$ ($3d^5$) have zero orbital angular
momentum. For HS-Mn$^{3+}$ ($3d^4$), a state of the $e_g$ symmetry is
unoccupied, also implying a lack of the orbital contribution to the
total magnetic moment of Mn$^{3+}$.

Thus, both $M(T)$ and $M(H)$ data point to a ferrimagnetic type of
ordering among the three magnetic sublattices of GdBaMn$_{2}$O$_{5.0}$.
Neutron powder diffraction studies \cite{MillangeY,MillangeLa} as well
as density-functional calculations of electronic structure
\cite{VidyaY,VidyaLa} have reported a rock-salt arrangement of Mn$^{2+}$
and Mn$^{3+}$ with a G-type AF ordering in YBaMn$_{2}$O$_{5.0}$ and
LaBaMn$_{2}$O$_{5.0}$, compounds with rare-earth ions from opposite ends
of the lanthanide series. It would be natural to assume that the same
types of charge and magnetic ordering among Mn ions are realized in
GdBaMn$_{2}$O$_{5.0}$. Magnetic Gd$^{3+}$ ions with a half-filled $4f$
shell are expected to interact ferromagnetically with Mn$^{3+}$ and
antiferromagnetically with Mn$^{2+}$ sublattices according to the
Goodenough-Kanamori rules \cite{G-K}, although this interaction should
be rather weak.

The magnetic structure most likely realized in GdBaMn$_{2}$O$_{5.0}$ is
shown in Fig. 2(b). Below $T_N$, Mn$^{2+}$ ($S$=5/2) and Mn$^{3+}$
($S$=2) ions develop a long-range ferrimagnetic order with the spin
directions along the $c$ axis. Because of the weak magnetic interaction
between Gd$^{3+}$ and Mn$^{2+}$/Mn$^{3+}$, the alignment of Gd spins
($S$=7/2) grows rather slowly with the development of the magnetization
of the Mn sublattices below $T_N$, but eventually almost all Gd spins
are aligned along the Mn$^{3+}$ spins at low enough $T$ and add to the
large $M_{sat}$. As can be seen in Fig 2(a), the magnetization vectors
of the ordered sublattices can be 90$^{\circ}$ rotated by applying a
magnetic field of $\sim$50 kOe perpendicular to the easy axis.

\begin{figure}
\includegraphics*[width=18pc]{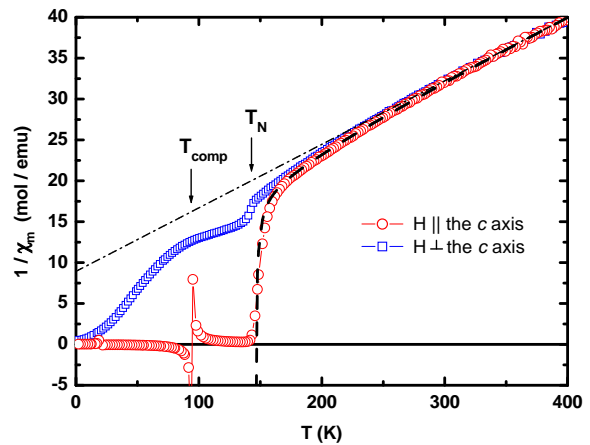}
\caption{(Color online) 
Inverse molar susceptibility $1/\chi_{m}$, measured in $H$ = 100 Oe
parallel (circles) and perpendicular (squares) to the $c$ axis, is in
agreement with a ferrimagnetic type of ordering. The dashed line is a
calculation within the molecular field theory (see text). The
dash-dotted line is a guide to the eye.}
\label{fig3}
\end{figure}

Another piece of information about the magnetic behavior of
GdBaMn$_{2}$O$_{5.0}$ comes from the susceptibility measurements in the
paramagnetic phase above $T_N$. Figure 3 shows the inverse molar
susceptibility, measured in $H$ = 100 Oe parallel (circles) and
perpendicular (squares) to the $c$ axis. Both curves coincide at high
temperature, demonstrating the isotropic nature of the paramagnetic
phase. In the molecular field theory \cite{Neel}, the temperature
dependence of the inverse susceptibility of a three-sublattice
ferrimagnetic material above $T_N$ is given by
\begin{equation}
\frac{1}{\chi} = \frac{1}{\chi_0} + \frac{T}{C}-\frac{\sigma(T)}{T-T_N},
\end{equation}
with $C$ the effective Curie constant, $T_N$ the N\'eel temperature, 
and $\sigma(T)=(\sigma_{0}T+m)/(T+\theta)$; $\chi_0$, $\sigma_{0}$,
$m$, and $\theta$ are all constants. The slope of the high-temperature
inverse susceptibility shown by the straight dash-dotted line in Fig. 3
is solely given by the effective Curie constant $C$, which is the sum of
the Curie constants of the three magnetic sublattices. The obtained
value of 15.1 emu\,K/mole is only 1\% less than what is expected for
Mn$^{3+}$ ($S$=2), Mn$^{2+}$ ($S$=5/2), and Gd$^{3+}$ ($S$=7/2)
\cite{note1}, giving further support to the proposed magnetic structure.

When one assumes that only Mn sublattices develop a long-range magnetic
order at $T_N$, the exchange coupling constant $J$ can be estimated, to
first approximation, from the following equation \cite{Kubo}:
\begin{equation}
T_N = \frac{q\,(2+\alpha+\beta)\,C_{\rm Mn^{3+}}C_{\rm Mn^{2+}}}
{C_{\rm Mn^{3+}}+C_{\rm Mn^{2+}}},
\end{equation}
where $C_{\rm Mn^{3+}}$ and $C_{\rm Mn^{2+}}$ are the Curie constants
for Mn$^{3+}$ and Mn$^{2+}$. Here, $q=(zJ)/(N_a g^2\mu_B^2)$ is the
molecular field constant related to the Mn$^{3+}$-Mn$^{2+}$ exchange
interaction, where $z$ = 6 is the number of nearest Mn$^{2+}$ neighbors
of Mn$^{3+}$ (and vice-versa), $N_a$ is Avogadro number, and the
coefficients $\alpha$ and $\beta$ give the strengths of the 
nearest-neighbor exchange interactions in each sublattice of
Mn$^{3+}$ and Mn$^{2+}$, respectively. Note that all constants in Eq.
(2) are determined entirely by these exchange interaction parameters and
the Curie constants. The fitting curve shown by the dashed line in Fig.
3 is calculated with $J/k_B$ = 30.5 K, $\alpha = -0.63$, and $\beta =
-0.69$, and it obviously reproduces the data very well, supporting the
assumption that Gd spins remain essentially paramagnetic across $T_N$.

\begin{figure}
\includegraphics*[width=20pc]{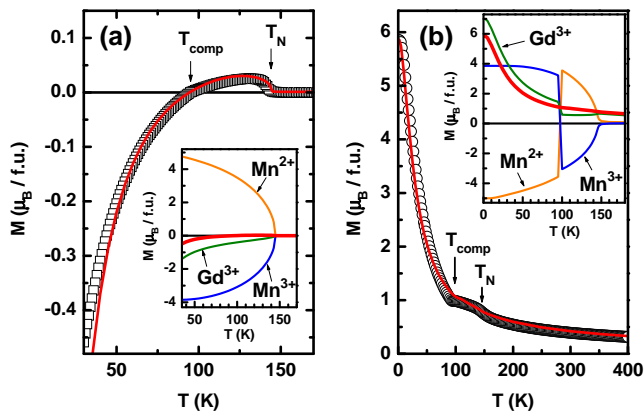}
\caption{(Color online) 
Temperature dependence of the easy-axis magnetization in (a) $H$ = 10 Oe
and (b) $H$ = 70 kOe. Solid lines are calculations within the molecular
field theory with the same parameters as for Fig. 3 (see text). Insets
show the contributions of individual sublattices to the overall
magnetization.}
\label{fig4}
\end{figure}

With the obtained exchange interaction parameters, the molecular field
theory \cite{Neel,Kubo} can be applied to the analysis of the
temperature dependence of $M$ in the ordered state below $T_N$. The
temperature-induced magnetization reversal observed at low magnetic
fields can be reproduced [as shown by the solid line in Fig. 4(a)] only
if there is a weak FM (AF) coupling between Gd$^{3+}$ ions and the
Mn$^{3+}$ (Mn$^{2+}$) sublattice. Furthermore, the value of the
compensation temperature is very sensitive to the strength of this
interaction. For $T_{\rm comp}$ = 95 K, the molecular field constant
related to the Gd$^{3+}$-Mn$^{3+}$ FM exchange is found to be about
0.01$q$, i.e., two orders of magnitude smaller than the AF exchange
interaction between Mn$^{3+}$ and Mn$^{2+}$. This value corresponds to
an effective magnetic field of approximately 35 kOe, while for the
ferrimagnetic spin order in the Mn$^{3+}$/Mn$^{2+}$ sublattices it is
$\sim$5000 kOe.

The large difference in the exchange interactions between
Mn$^{3+}$-Mn$^{2+}$ and Gd-Mn is crucial for the understanding of the
observed magnetic behavior in GdBaMn$_{2}$O$_{5.0}$. Just below $T_N$,
Mn$^{3+}$ spins, being antiferromagnetically coupled to the Mn$^{2+}$
sublattice, are opposite to the applied magnetic field, and at low
fields Gd spins tend to align in the same direction as the Mn$^{3+}$
sublattice [see inset of Fig. 4(a)], though their alignment is weak;
below $T_{\rm comp}$, the sum of the magnetic moments of Gd$^{3+}$ and
Mn$^{3+}$ overgrows the magnetic moment of Mn$^{2+}$, giving rise to a
negative magnetization. Intriguingly, the situation changes completely
in a high magnetic field which is strong enough to compete with the
Gd-Mn exchange interaction: In this case, Gd spins tend to align along
the external magnetic-field direction, causing a positive magnetization
at all temperature as shown in Fig. 4(b), and the magnetization reversal
is eliminated. It is noteworthy that the Mn spins also behave
differently and shows a novel turnabout; namely, just below $T_N$ their
orientation is the same as in the low magnetic field case, but with
decreasing temperature this state becomes energetically unfavorable as
the influence of Gd spins grows. As a result, near the compensation
point, an abrupt turnabout of the magnetization of the Mn sublattices
takes place [see inset of Fig. 4(b)], which is manifested in the sharp
change of the slope of $M(T)$. Note that in the present case the strong
Ising anisotropy probably plays a key role in the abrupt turnabout.

To conclude, the present study shows that unusual magnetic behavior of a
new three-sublattice ferrimagnetic manganite GdBaMn$_{2}$O$_{5.0}$ --- a
temperature-induced magnetization reversal at low magnetic fields and a
novel turnabout of Mn sublattice magnetizations at high magnetic fields
--- are governed by an elaborate interplay between its magnetic
sublattices that are formed by the $A$-site ordering as well as the
charge ordering of Mn into Mn$^{2+}$ and Mn$^{3+}$. In the present case,
the weak coupling of Gd spins with magnetically ordered
Mn$^{2+}$/Mn$^{3+}$ sublattices is the source of novel ferrimagnetism
not found elsewhere before. This result not only enriches our knowledge
about ferrimagnetics, but also points to the potential of
charge-order-susceptible transition-metal oxides for discovering
physically interesting and technologically useful properties.

\begin{acknowledgments}
We thank I. Tsukada for helpful discussions.
\end{acknowledgments}


\begin{thebibliography}{99}
\bibitem{note1} The Curie constant $C_i$ of an individual sublattice is
$C_i=N_a (g \mu_B)^2 S_i(S_i+1)/3k_B$,
where $S_i$ is the spin of the magnetic ion $i$.
\end{thebibliography}
\end{document}